\documentclass[aip,jap,a4,reprint,twoside,floatfix,superscriptaddress]{revtex4-1} 

\usepackage{epsfig}                                                            
\usepackage{graphicx}
\usepackage{dcolumn}
\usepackage{color}
\usepackage{bm}
\usepackage{subfigure} 
\usepackage{epsfig} 

\begin{document}

\title {\bf Large magnetocapacitance in electronic ferroelectric manganite systems }

\author{Ujjal Chowdhury}
\affiliation{Nanostructured Materials Division, CSIR-Central Glass and Ceramic Research Institute, Kolkata 700032, India}
\author{Sudipta Goswami}
\affiliation{Nanostructured Materials Division, CSIR-Central Glass and Ceramic Research Institute, Kolkata 700032, India}
\author{Dipten Bhattacharya} \email{dipten@cgcri.res.in}
\affiliation{Nanostructured Materials Division, CSIR-Central Glass and Ceramic Research Institute, Kolkata 700032, India}
\author{Arindam Midya}
\affiliation{Experimental Condensed Matter Physics, Saha Institute of Nuclear Physics, Kolkata 700064, India }
\author{P. Mandal}
\affiliation{Experimental Condensed Matter Physics, Saha Institute of Nuclear Physics, Kolkata 700064, India }
\author{Pintu Das}
\affiliation{Institute of Physics, J.W. Goethe University, D-60438 Frankfurt, Germany}
\author{Ya. M. Mukovskii}
\affiliation{National Research Technological University, "MISiS", Moscow 119049, Russia}

\date{\today}

\begin{abstract}
 
We have observed a sizable positive magnetocapacitance ($\sim$5-90\%) in perovskite Pr$_{0.55}$Ca$_{0.45}$MnO$_3$ and bilayer Pr(Sr$_{0.1}$Ca$_{0.9}$)$_2$Mn$_2$O$_7$ system under 5T magnetic field across 20-100 K below the magnetic transition point T$_N$. The magnetodielectric effect, on the other hand, exhibits a crossover: (a) from positive to negative for the perovskite system and (b) from negative to positive for the bilayer system over the same temperature range. The bilayer Pr(Sr$_{0.1}$Ca$_{0.9}$)$_2$Mn$_2$O$_7$ system exhibits a sizable anisotropy as well. We have also noticed the influence of magnetic field on the dielectric relaxation characteristics of these systems. These systems belong to a class of improper ferroelectrics and are expected to exhibit charge/orbital order driven ferroelectric polarization below the transition point T$_{CO}$. Large magnetocapacitance in these systems shows typical multiferroic behavior even though the ferroelectric polarization is small in comparison to that of other ferroelectrics. 

\end{abstract}

\maketitle

\section{Introduction}

A series of improper ferroelectrics - geometric, magnetic, and electronic - has been found to exhibit the multiferroic coupling as well.\cite{Cheong, Stroppa} The extent of coupling, however, varies. Among them, the magnetic ferroelectrics (TbMnO$_3$, DyMnO$_3$) exhibit stronger coupling though at a lower temperature (below the commensurate to incommensurate magnetic transition temperature $T_{IC}$ $\sim$28 K). The magnetic transition point for the geometric ferroelectrics (YMnO$_3$, LuMnO$_3$) too turn out to be low ($\sim$80-100 K). For both of these classes, detailed work has already been done to reveal the magnitude of the coupling as well as its underlying mechanism.\cite{Khomskii} Except LuFe$_2$O$_4$ [Ref. 4], the electronic ferroelectrics, on the other hand, seem to have received relatively lesser attention in spite of the fact that in quite a few of them both the ferroelectric and magnetic transition points are higher and closer to room temperature. The electronic ferroelectricity originates from noncentrosymmetric charge/orbital order\cite{van} - even frustrated charge order - and is expected to offer a sizable polarization ($\sim$5-6 $\mu$C/cm$^2$).\cite{Dagotto,Picozzi} However, because of intrinsic electronic inhomogeneity and consequent large leakage current, it is difficult to measure the polarization directly by global electrical measurements.\cite{Lopes} 

In several other compounds too, such as, Pr$_{1-x}$Ca$_x$MnO$_3$ (x = 0.45-0.55)\cite{Efremov}, Fe$_3$O$_4$,\cite{Yamauchi} Pr(Sr$_{0.1}$Ca$_{0.9}$)$_2$Mn$_2$O$_7$ etc,\cite{Tokunaga} broken symmetry of the charge/orbital order creates polar domains. Yet the extent of multiferroic coupling in them has not been investigated in detail. It is important to mention here that even in a well-known multiferroic compound TbMnO$_3$, which exhibits very strong coupling, the ferroelectric polarization is small ($\sim$0.09 $\mu$C/cm$^2$).\cite{Goto} Therefore, small polarization in electronic ferroelectrics need not be detrimental to strong multiferroicity. Motivated by this observation and also because of relatively higher ferroelectric and magnetic transition points in electronic ferroelectric systems, we explored, in this work, the extent of magnetocapacitance in two such compounds - bilayer Pr(Sr$_{0.1}$Ca$_{0.9}$)$_2$Mn$_2$O$_7$ and perovskite Pr$_{0.55}$Ca$_{0.45}$MnO$_3$. The bilayer system undergoes two consecutive charge/orbital order transitions at $T_{CO1}$ $\sim$380 K and at $T_{CO2}$ $\sim$300 K. The crystallographic structure transforms from orthorhombic Amam to Pbnm at $T_{CO1}$ and from Pbnm to Am2m at $T_{CO2}$ with the rotation of orbital ordered chains from along b- to a-axis. The chains stack in the form of stripes along a- and b-axis, respectively, in these phases. This rotation is associated with a rearrangement of Mn$^{3+}$ and Mn$^{4+}$ ions to yield a polarization along b-axis.\cite{Tokunaga} The three-dimensional magnetic structure, sets in at $T_N$ $\sim$150 K, is antiferromagnetic with easy axis oriented along b-axis. In the case of the perovskite system, on the other hand, mixed site- and bond-centered charge ordered structure breaks the symmetry and yields a polarization at the charge order transition point $\sim$220 K.\cite{van} For this system the magnetic transition point T$_N$ is $\sim$100 K.\cite{Dediu} We observed a sizable magnetocapacitance ($\sim$5-90\%) under 5T field, in these systems, across 20-100 K below the magnetic transition point (T$_N$). 

\section{Experiments}
The single crystals of Pr(Sr$_{0.1}$Ca$_{0.9}$)$_2$Mn$_2$O$_7$ and Pr$_{0.55}$Ca$_{0.45}$MnO$_3$ have been grown by float-zone technique in an image furnace.\cite{Mukovskii} The disc shaped samples of diameter $\sim$2.0-3.0 mm and thickness $\sim$0.5-1.0 mm have been extracted from the boule. The Laue pattern and EDX data have been included in the supplementary document.\cite{supplemental} Fresh surface has been cleaved out for all the electrical measurements. The measurements have been carried out in two-probe configuration using silver conducting paste and silver wires. Apart from dc resistivity, dielectric spectra have been recorded across 100 Hz - 5 MHz with 1V amplitude over a temperature range 20-100 K both under zero and 5T magnetic field. A Cryogenic Instrument cryostat and a superconducting magnet have been used for covering the temperature and field range while an impedance analyzer of Wayne Kerr (Model 6500B) has been used for the dielectric property measurement. The protocol of dielectric measurements was as follows. The sample was cooled down from room temperature under zero magnetic field to 20 K and the dielectric spectra were recorded at several temperatures in between 20 and 100 K. The sample was heated back to the room temperature and then cooled down again to 20 K under zero field. A magnetic field of 5T was then applied and the dielectric spectra were recorded again at those temperatures in between 20 and 100 K. We restricted all the dielectric measurements within a temperature regime of 20-100 K as with the rise in temperature the conductivity of the sample increases. This, in turn, makes accurate determination of the intrinsic capacitance of the samples difficult as contribution from interface starts dominating the overall dielectric response. Since our main aim in this paper was to determine the extent of magnetocapacitance in these electronic ferroelectric manganite systems, we carried out all the measurements within a temperature range where the influence of conductivity is small. 

\section{Results and Discussion}
In Fig. 1 we show the dc resistivity ($\rho$) versus temperature ($T$) patterns for the Pr(Sr$_{0.1}$Ca$_{0.9}$)$_2$Mn$_2$O$_7$ (PSCMO) and Pr$_{0.55}$Ca$_{0.45}$MnO$_3$ (PCMO) samples across the temperature range of our interest. The $\rho-T$ has been measured along parallel ($\rho_c$) and perpendicular to the c-axis ($\rho_{ab}$). For the bilayer sample, reasonably large anisotropy could be observed whereas for the perovskite one, not much anisotropy has been observed. Since not much anisotropy could be observed, all the data presented here for the perovskite sample have been obtained from measurements within the ab-plane. The onset of magnetic order is shown by a distinct anomaly around T$_N$ in all the $\rho-T$ plots. For the bilayer system, there is a hysteresis around T$_N$ in the $\rho_{ab}-T$ signifying first-order transition while $\rho_c-T$ does not show any hysteresis. The first order transition within ab-plane possibly results from stronger influence of magnetic order with easy axis oriented along b-axis on the charge/orbital order essentially confined within ab-plane and stacked along c-axis. A second hysteresis could also be seen in the $\rho_{ab}-T$ plot around 110 K which possibly reflects the low temperature structural transition (from orthorhomic to monoclinic) observed by others.\cite{Tokunaga,Beale} For the perosvkite system, of course, no such hysteresis in $\rho-T$ pattern could be observed around T$_N$ (inset of Fig. 1b).

\begin{figure}[!ht]
  \begin{center}
   \subfigure[]{\includegraphics[scale=0.18]{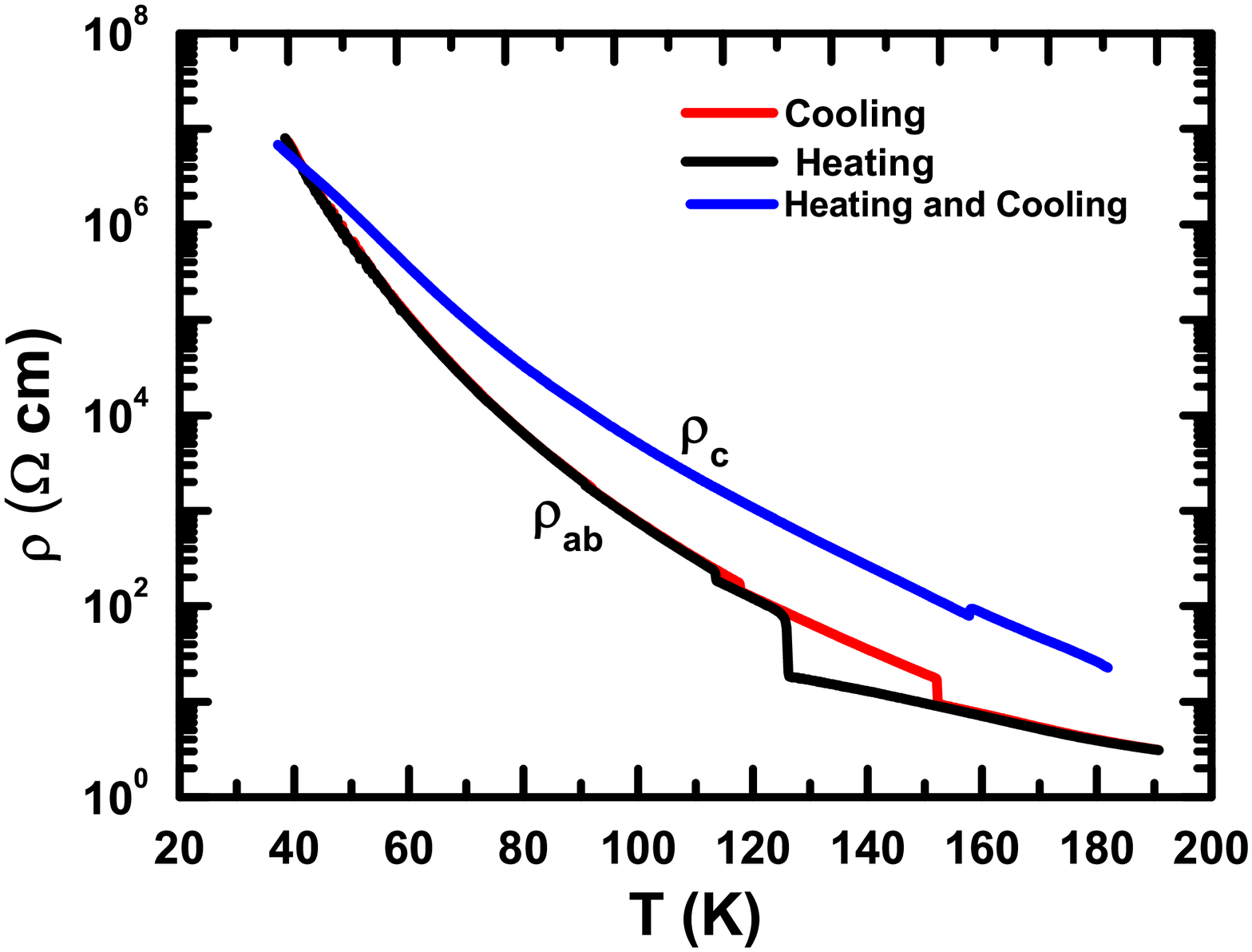}}
    \subfigure[]{\includegraphics[scale=0.18]{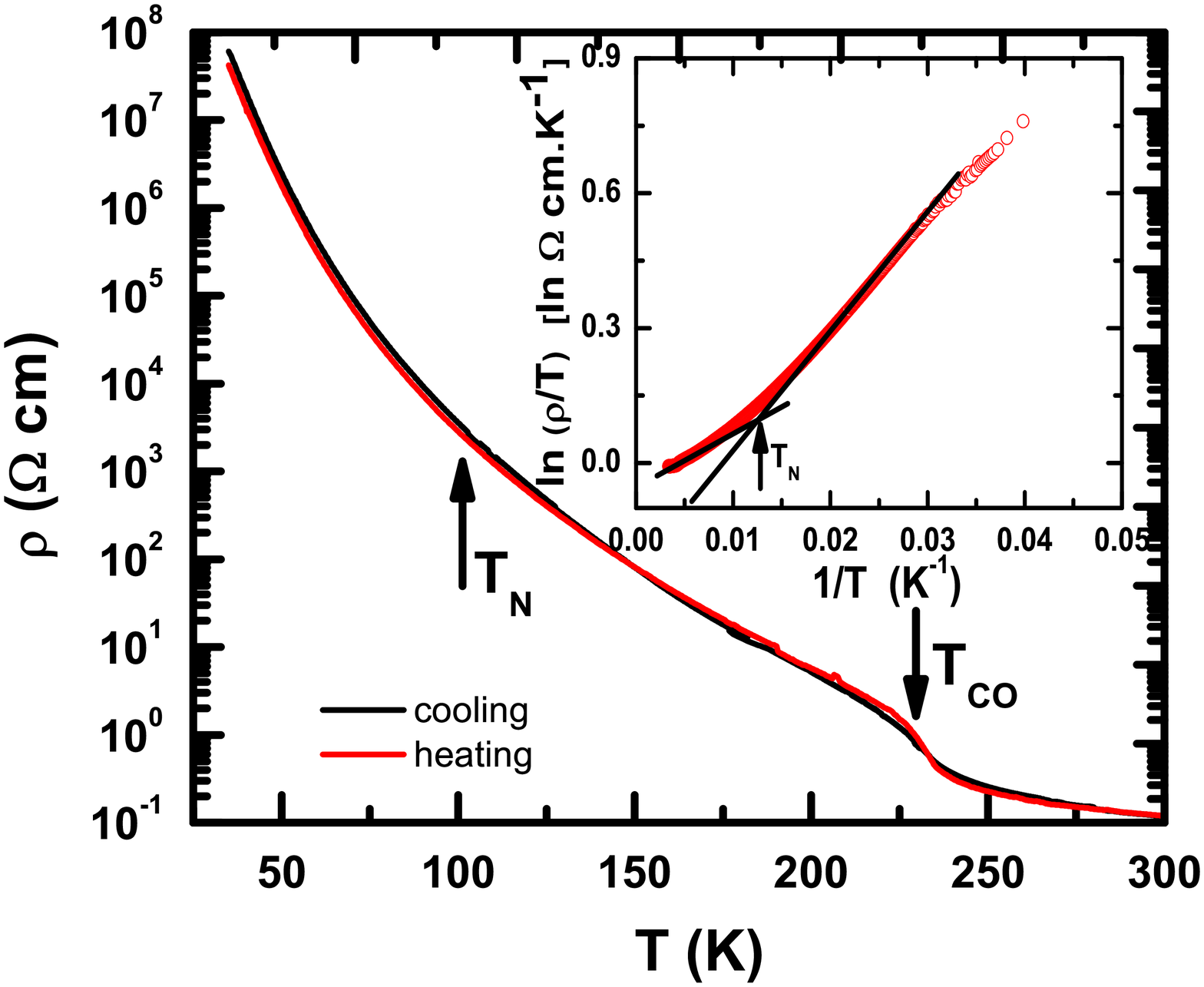}}  
    \end{center}
  \caption{(color online) The resistivity versus temperature plots showing the anomaly around T$_N$ for both the (a) bilayer (along and perpendicular to the c-axis) and the (b) perovskite systems; inset shows the anomaly around T$_N$. }
\end{figure}

\begin{figure*}[!htp]
  \begin{center}
    \subfigure[]{\includegraphics[scale=0.20]{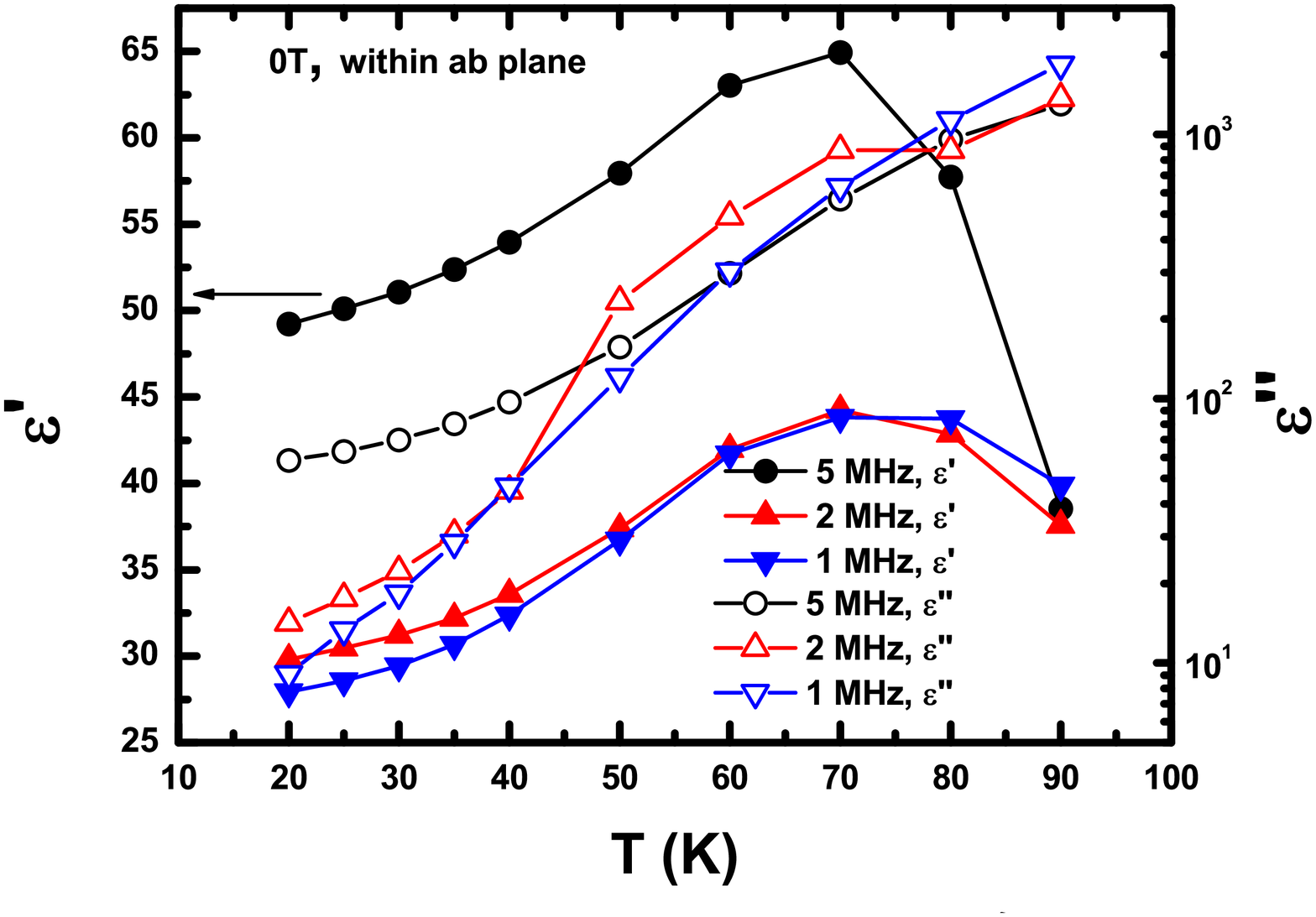}}
    \subfigure[]{\includegraphics[scale=0.27]{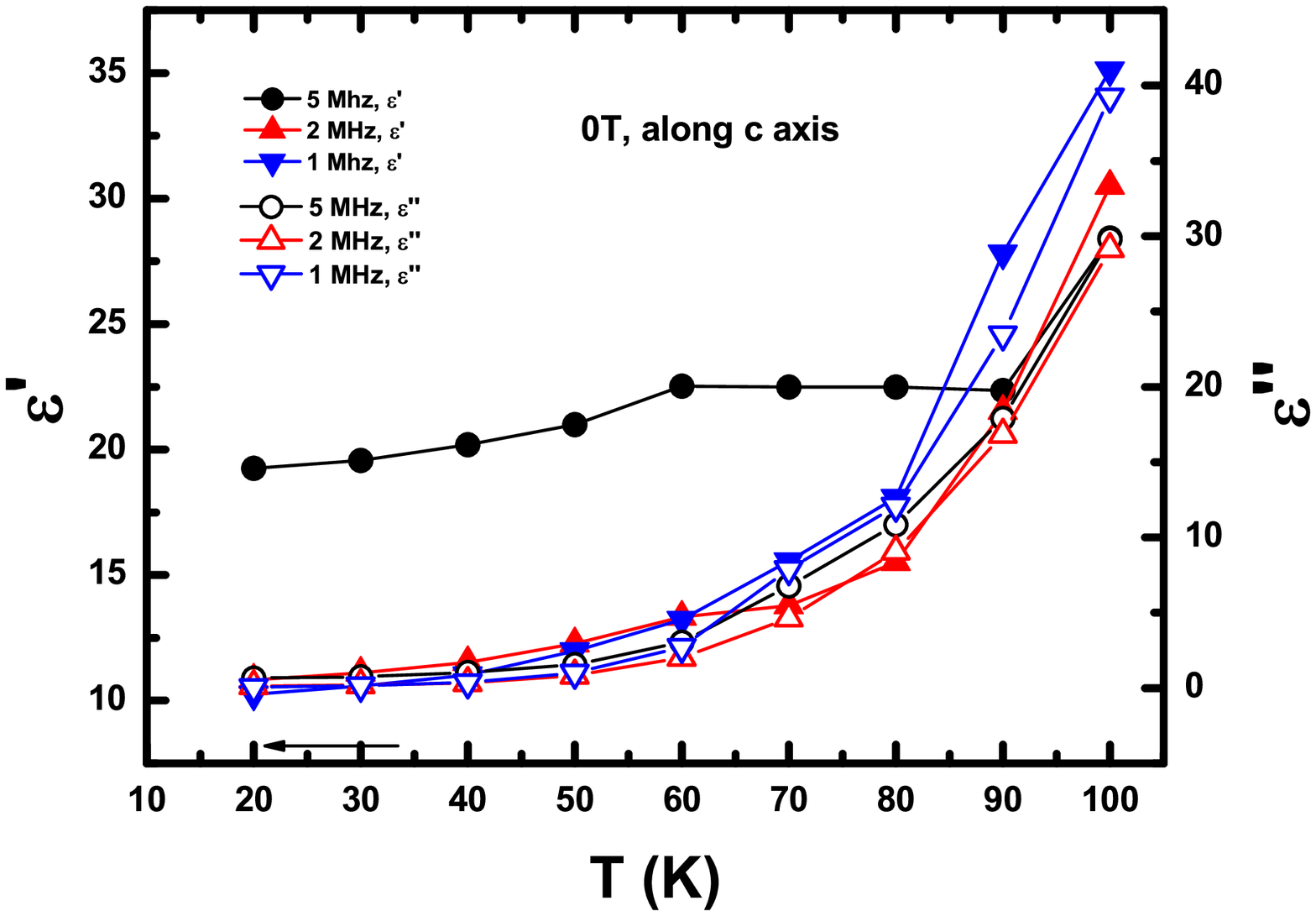}}  
	 \subfigure[]{\includegraphics[scale=0.27]{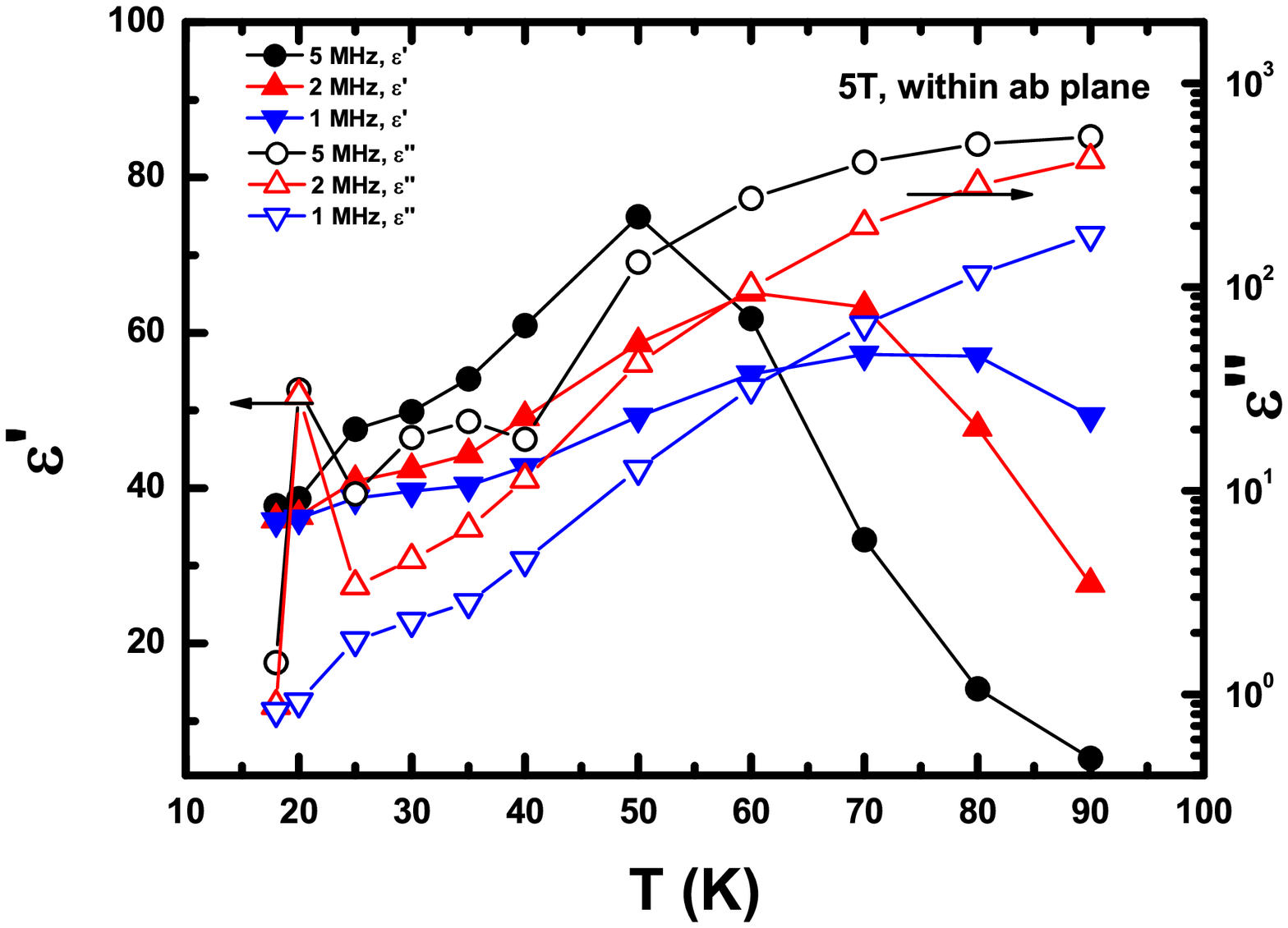}}
    \subfigure[]{\includegraphics[scale=0.27]{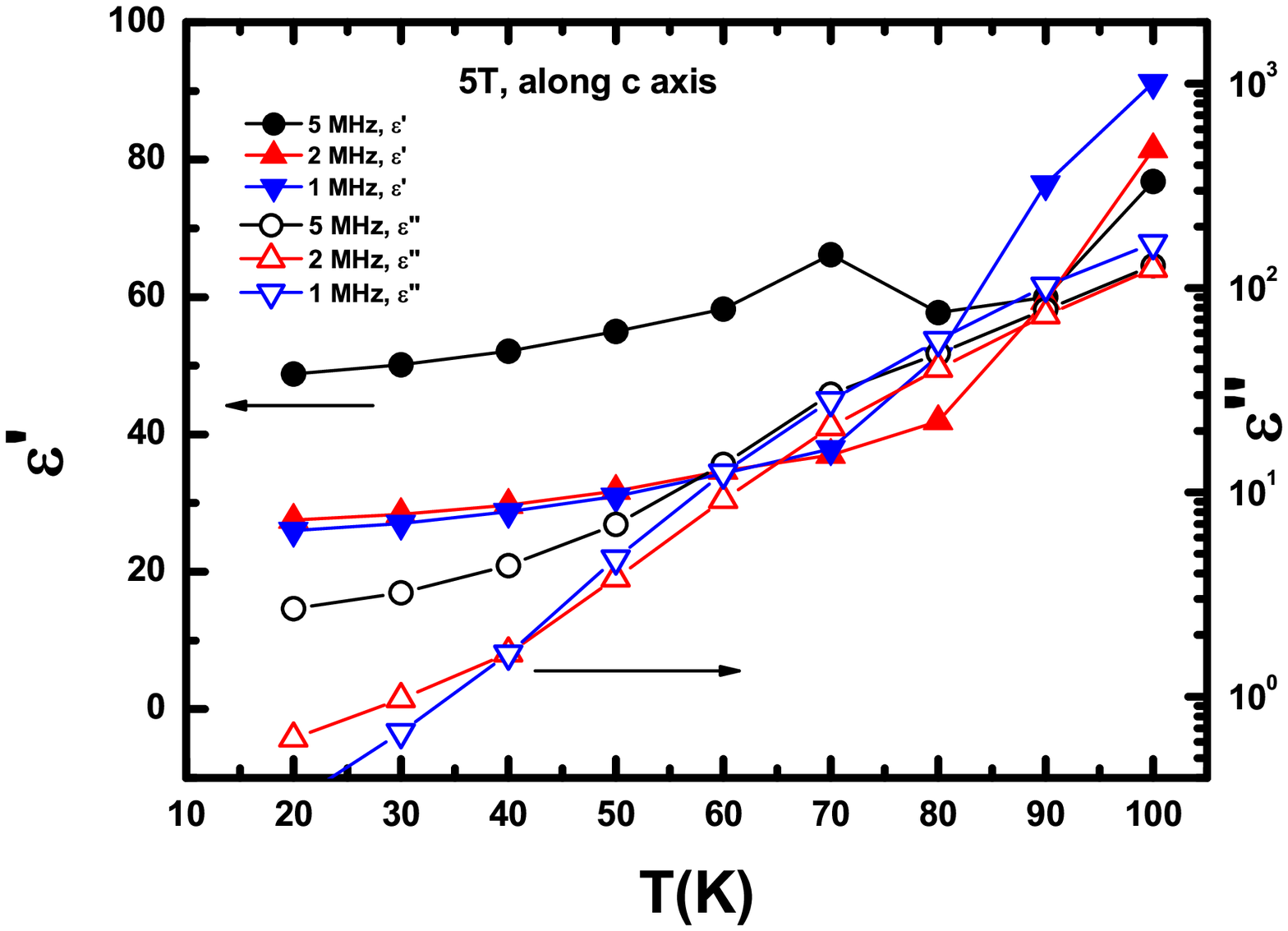}} 
	 \subfigure[]{\includegraphics[scale=0.19]{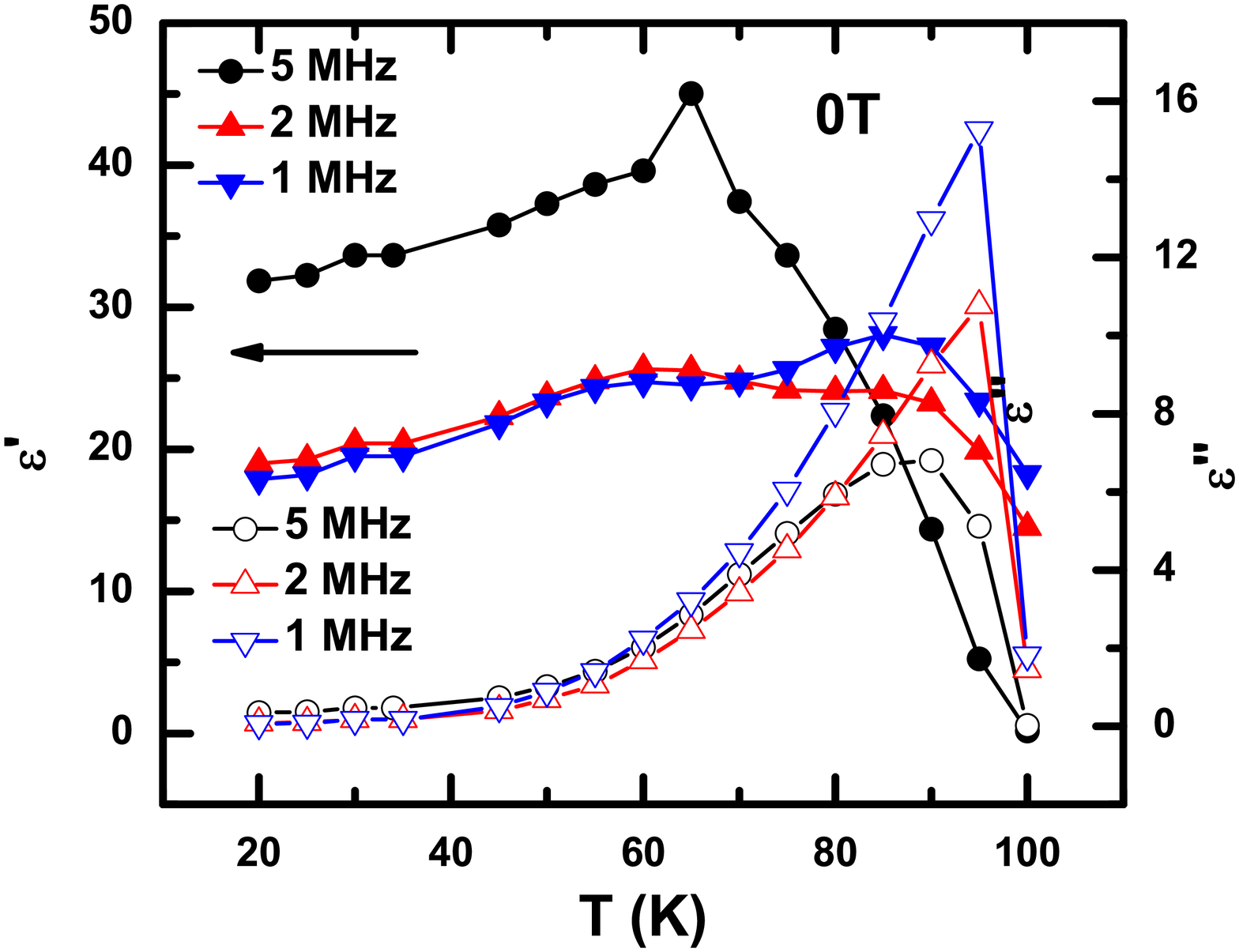}}
    \subfigure[]{\includegraphics[scale=0.19]{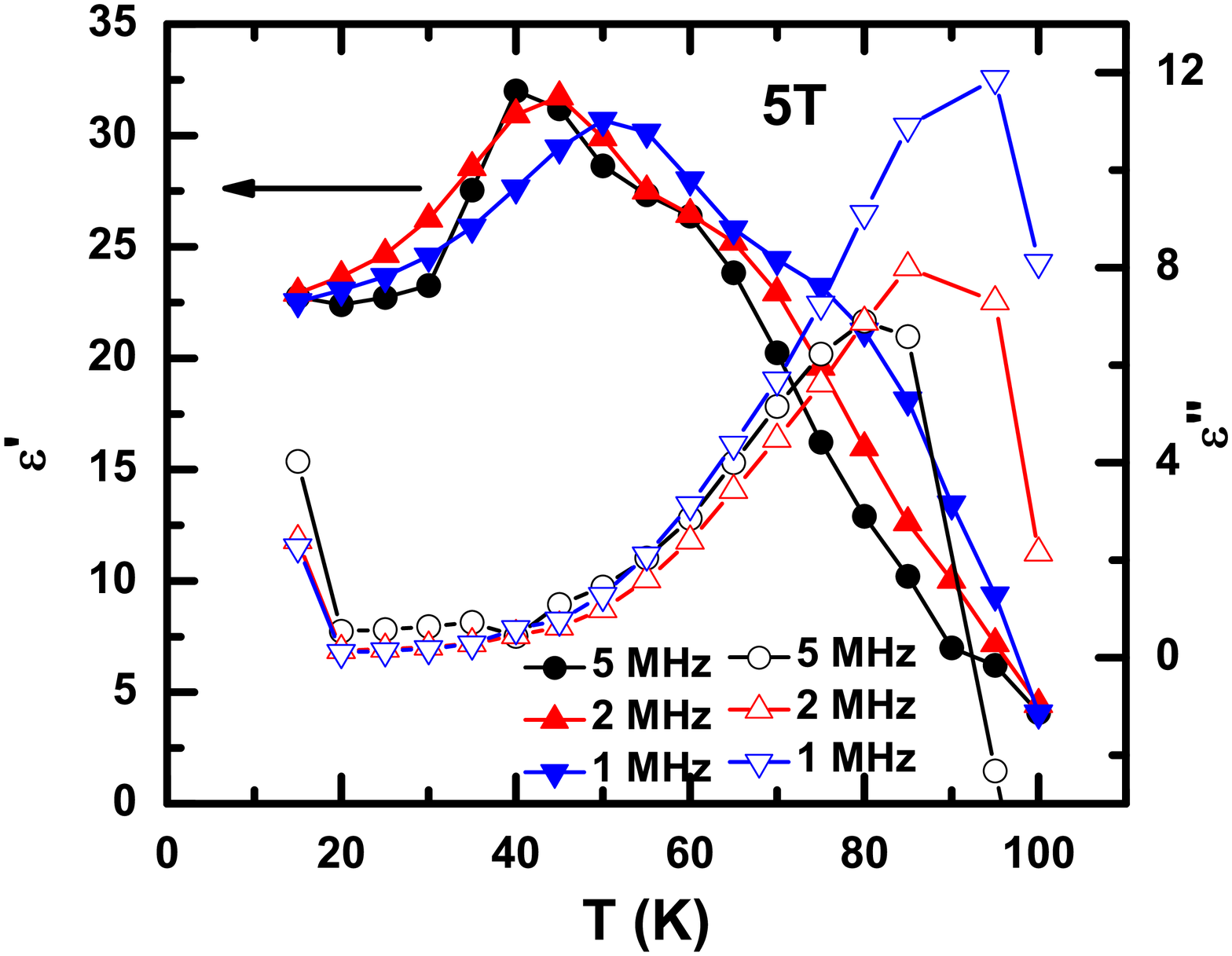}} 
  \end{center}
  \caption{(color online) The real ($\epsilon'$) and imaginary ($\epsilon"$) dielectric permittivity versus temperature plots for the bilayer system (a) within ab-plane and (b) along c-axis under zero field; and the corresponding plots (c) within ab-plane and (d) along c-axis under 5T field; (e) and (f) show the plots for the perovskite sample under zero and 5T field, respectively.}
\end{figure*}

In Fig. 2, we show the real and imaginary parts of the permittivity ($\epsilon'$, $\epsilon"$) across 20-100 K measured under zero and 5T field for both the samples. For the bilayer sample, the $\epsilon'_{ab}$, $\epsilon"_{ab}$ and $\epsilon'_c$, $\epsilon"_c$ versus temperature plots under zero field at different frequencies are shown in Figs. 2(a) and 2(b) whereas the corresponding plots under 5T field are shown in Figs. 2(c) and 2(d), respectively. For the perovskite sample, the plots under zero and 5T field are shown in Figs. 2(e) and 2(f), respectively. In this case, only the data obtained from the measurement within ab-plane are shown. Qualitatively all the plots exhibit nearly similar features. There is a maximum in the real part of the permittivity which shifts toward lower temperature under the magnetic field. This could be associated with a crossover in the mechanism of polarization in these samples. While at lower temperature the ferroelectric order governs the polarization, with the rise in temperature, enhanced contribution of the charge carrier hopping starts playing an important role. Signature of such a crossover could be observed in the plot of magnetocapacitance versus temperature as well (discussed later). 

Fig. 3 shows the \% change in the real part of the permittivity ($\frac{\epsilon'_{5T}-\epsilon'_0}{\epsilon'_{5T}}$ = $\Delta\epsilon'/\epsilon'$) at a frequency of 2 MHz under 5T field. For the bilayer PSCMO system, $\Delta\epsilon'/\epsilon'$ is negative and its magnitude is $\sim$20\% at low temperature (20-30 K). The magnitude increases up to $\sim$60\% around 50 K and then decreases and eventually changes the sign to become positive and reaches $\sim$30\% at 90 K. Interestingly, the temperature dependence of the magnetodielectric data measured both within and perpendicular to ab-plane nearly resemble each other. This is quite expected as the dielectric permitivity versus temperature plots too, do not show any significant change in the patterns (Fig. 2). There is a slight frequency dependence as well whose significance will be clear in the analyses that followed. The extent of anisotropy varies within 2-5 across the entire temperature range. In the case of the perovskite PCMO sample, on the other hand, the $\Delta\epsilon'/\epsilon'$-$T$ plot turns out to be quite different. Here, $\Delta\epsilon'/\epsilon'$ is positive and $\sim$40\% at low temperature (20-50 K). It decreases and changes sign at $\sim$75 K to become negative and then increases sharply up to $\sim$250\% by $\sim$100 K. Therefore, while the bilayer sample exhibits a switch from negative to positive, the perovskite one exhibits a switch from positive to negative as far as the magnetodielectric effect is concerned over a temperature range 20-100 K.

\begin{figure}[!ht]
  \begin{center}
    \includegraphics[scale=0.25]{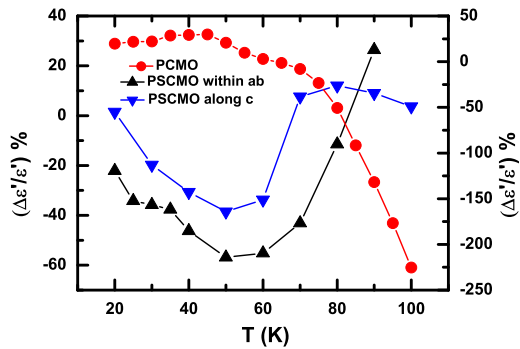} 
    \end{center}
  \caption{(color online) The magnetodielectric effect under 5T field for the bilayer (left scale) - both within ab-plane and along c-axis - and the perovskite (right scale) systems. }
\end{figure}

The large magnetodielectric effect reflects a strong multiferroic coupling in these systems. The origin of the multiferroic coupling in electronic ferroelectric systems could either be fundamental - the coupling between charge/orbital and spin degrees of freedom\cite{Kugel,Paolasini,Chuang} - or the striction effect as in other conventional multiferroic systems. The dielectric constant under a magnetic field, in the present cases, has both decreased or increased with respect to that under zero field at low temperature ($\sim$20 K) and then changed substantially as the temperature is raised in order to trigger a change in the sign of the magnetodielectric effect (Fig. 3). It has been found already that in different multiferroic systems such as BiFeO$_3$ [Ref. 20], LuFe$_2$O$_4$ [Ref. 21], LiVCuO$_4$ [Ref. 22], the polarization or dielectric constant is suppressed at the onset of magnetic order. In BiFeO$_3$, the suppression results from the structural effects - enhanced antiferrodistortive rotation of the octahedra around the axis of polarization at the onset of magnetic order leads to the suppression of the noncentrosymmetry.\cite{Goswami} There are, however, contradictory results as well. For instance, in LuFe$_2$O$_4$, while Ikeda $\textit{et al}$.\cite{Ikeda} have observed a rise in the polarization at T$_N$, Subramanian $\textit{et al}$.\cite{Subramanian} have noticed a drop in the dielectric constant under a magnetic field below T$_N$. Switching of the polarization axis under magnetic field\cite{Goto} - as noted in TbMnO$_3$ and DyMnO$_3$ - from along one to another crystallographic axis can, of course, lead to such a contradictory result when measured in a rather polycrystalline sample or in a twinned crystal because of variation in the averaging effect. It has been clearly shown that for strongly correlated electron systems, the charge/orbital order is intimately coupled with the spin order\cite{Paolasini,Chuang} and, in fact, spin order helps in strengthening the charge/orbital order and even recovering the charge order lost under photoexcitation. Therefore, in the present case, where the ferroelectricity is originating from the noncentrosymmetric charge/orbital order, it could be expected that the polarization or the dielectric constant would increase at the onset of magnetic order at T$_N$ or under a magnetic field at below T$_N$. However, we do not notice any consistent pattern of increase or suppression of the dielectric constant under the magnetic field for both the samples. We contrast this result with the result of magnetocapacitance (shown and discussed below) to point out that the dielectric constant, measured even under a relatively higher frequency (1-5 MHz), does not really represent the data corresponding to a relaxed state\cite{Loidl} and, therefore, could be misleading as far as the behavior of the intrinsic polarization is concerned. We also point out that we observe a wide dispersion of the dielectric permittivity with frequency and the relaxation dymanics deviates from the ideal Debye-type relaxation by a large extent signifying a much broader range of relaxation time constant ($\tau$). As a result, as mentioned above, the magnetodielectric effect (measured even at a higher frequency 1-5 MHz) exhibits a slight frequency dependence. 

\begin{figure*}[!htp]
   \begin{center}
    \subfigure[]{\includegraphics[scale=0.25]{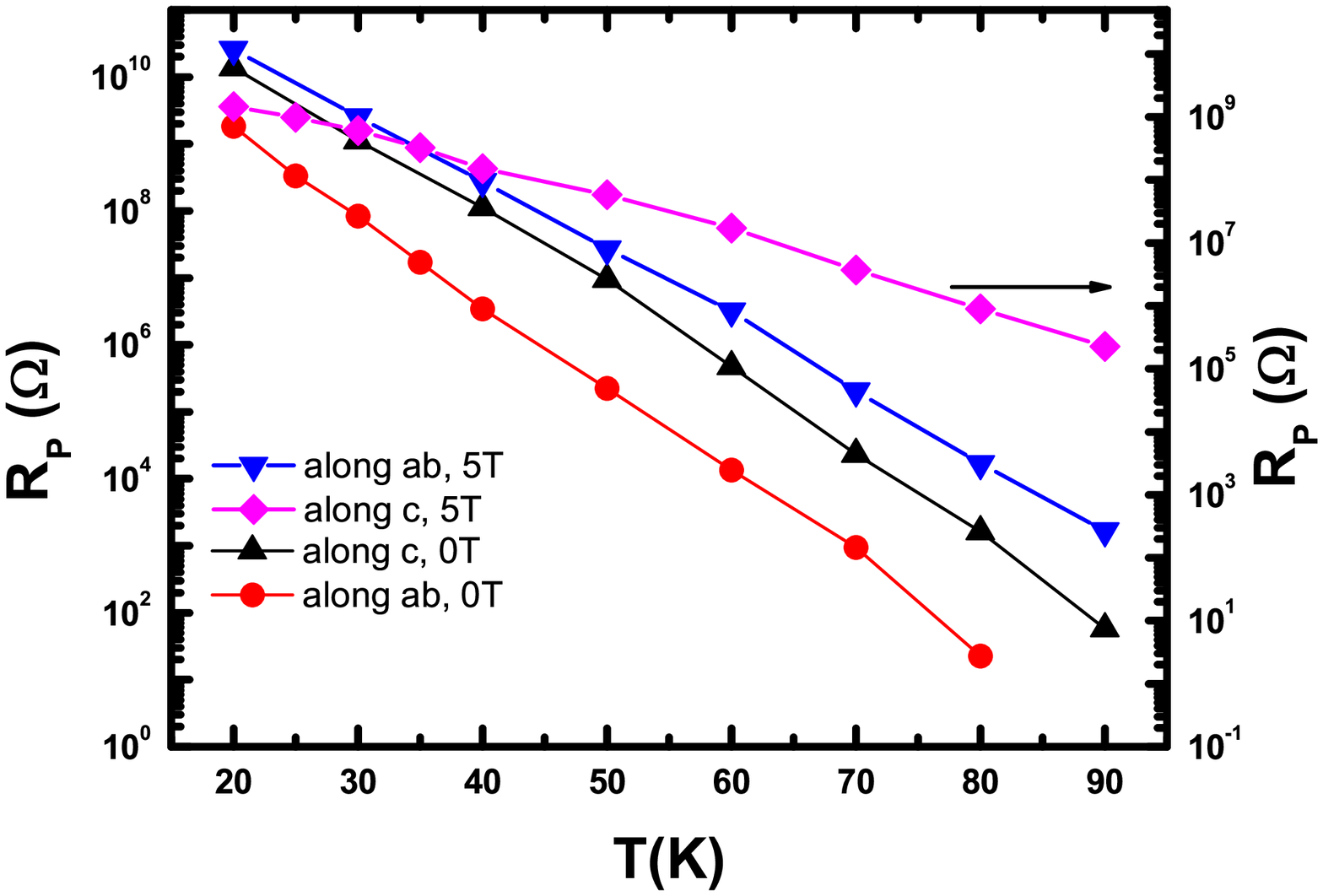}}
    \subfigure[]{\includegraphics[scale=0.25]{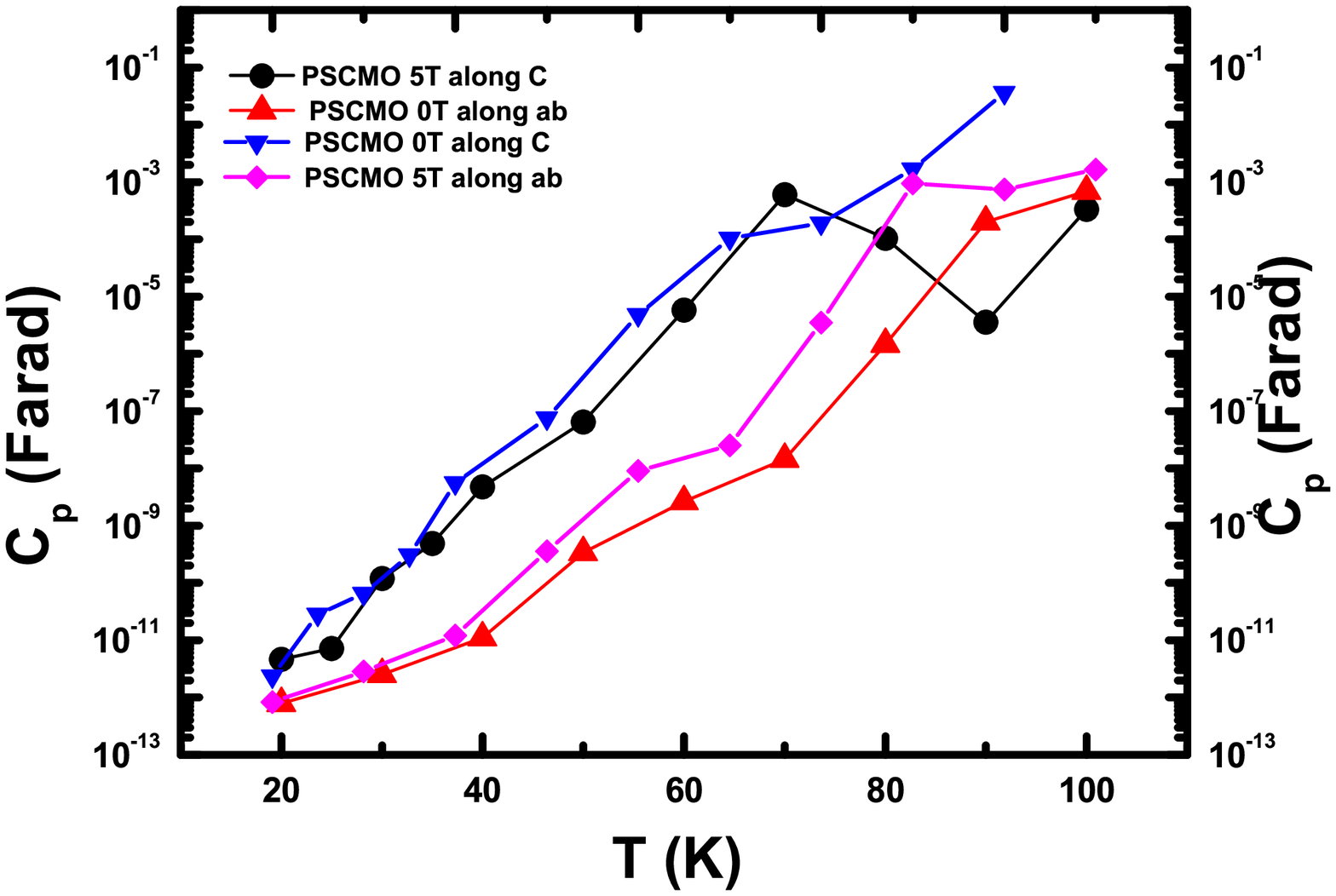}}  
	 \subfigure[]{\includegraphics[scale=0.19]{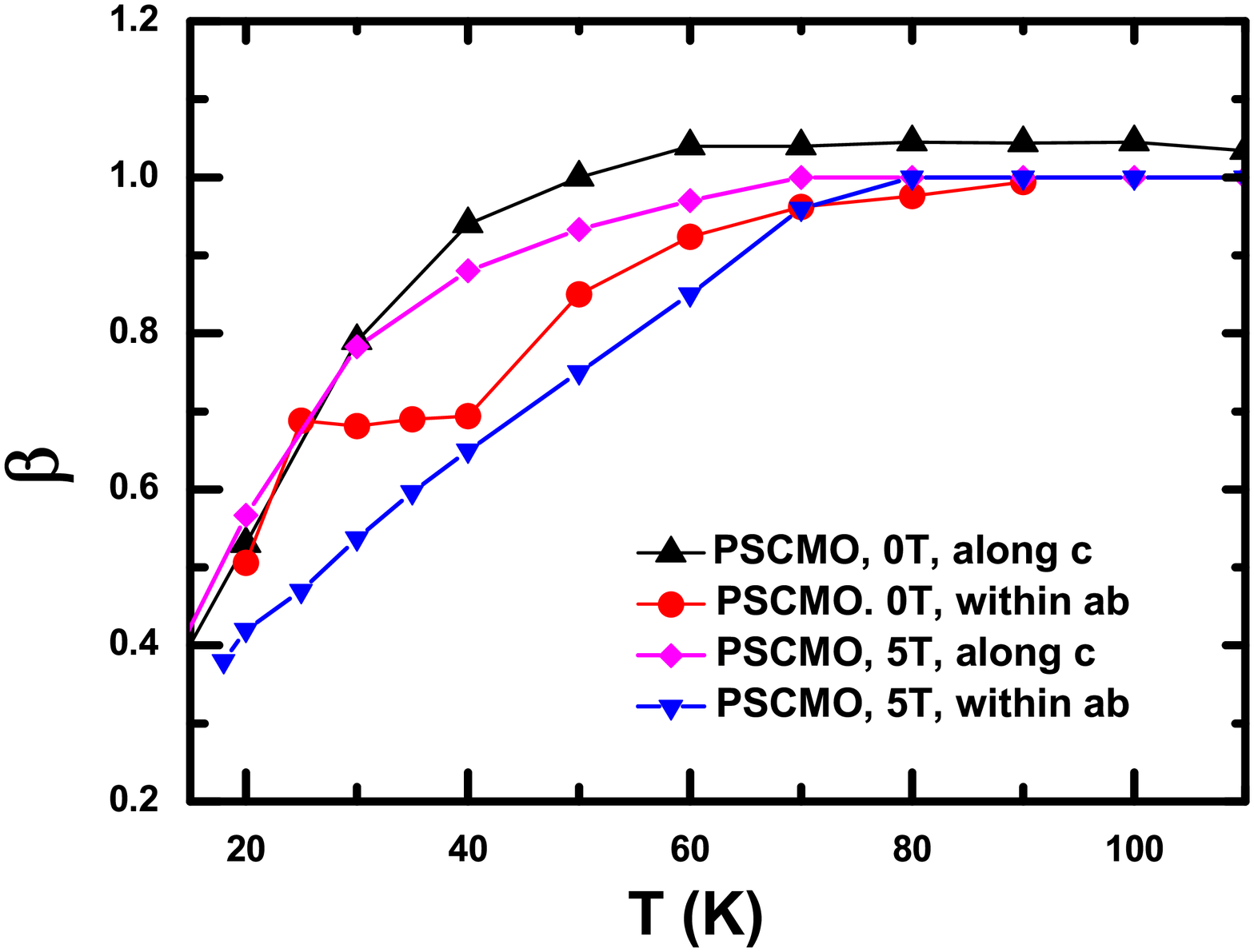}}
    \subfigure[]{\includegraphics[scale=0.19]{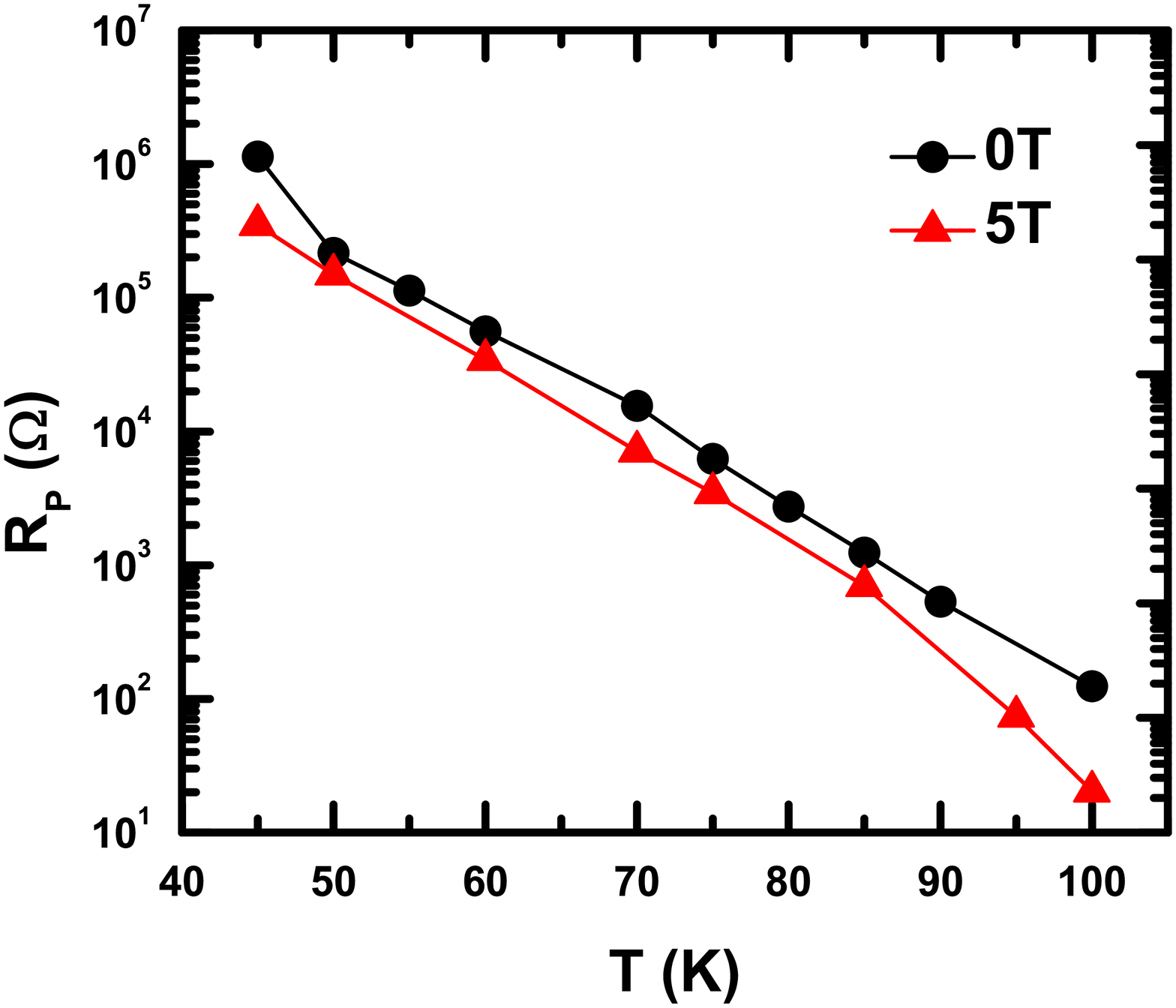}} 
	 \subfigure[]{\includegraphics[scale=0.19]{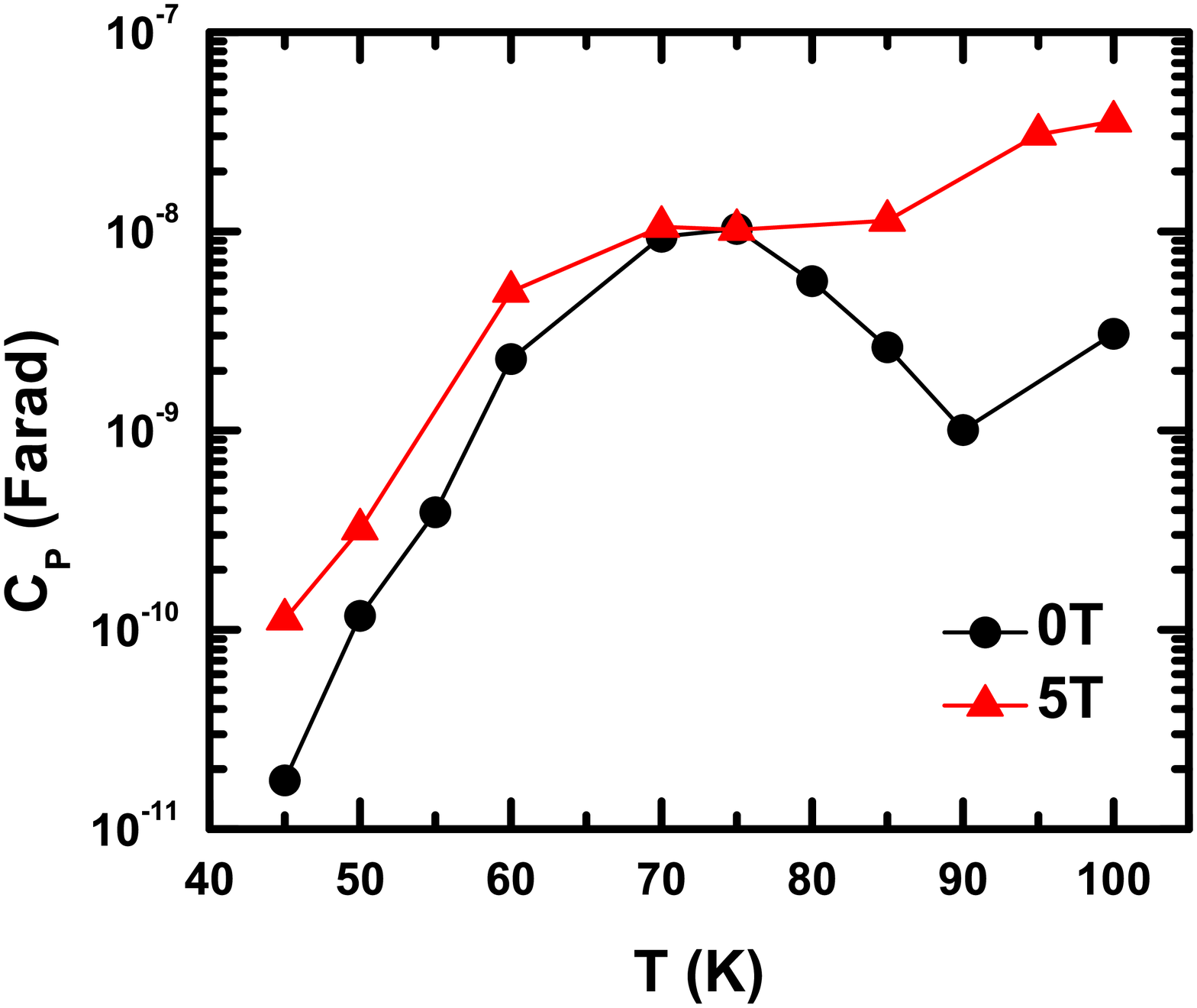}}
    \subfigure[]{\includegraphics[scale=0.19]{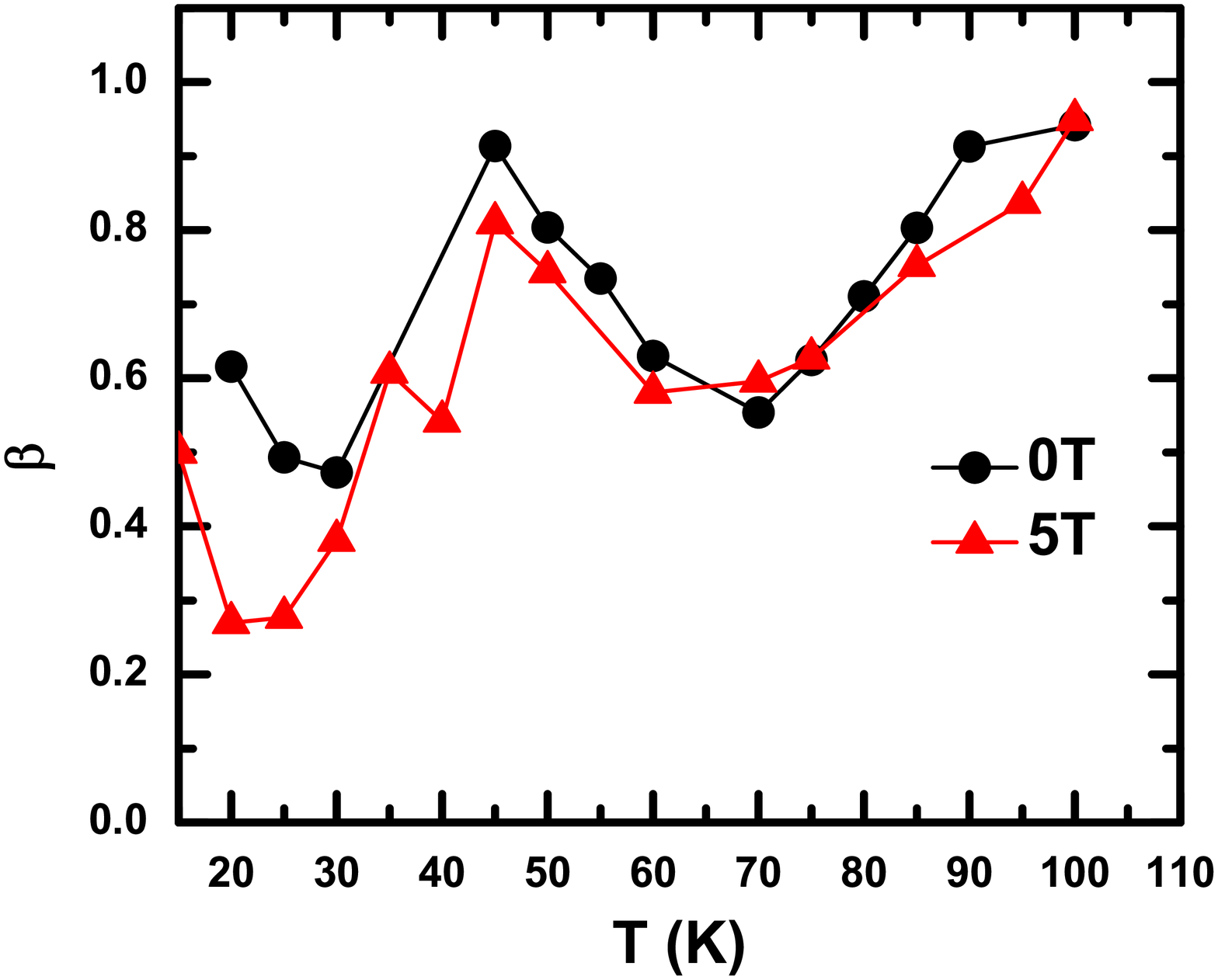}} 
  \end{center}
  \caption{(color online) The circuit parameters (a) R$_p$, (b) C$_p$, and (c) $\beta$ as a function of temperature - both along c-axis and within ab-plane - determined from the fitting of the dielectric relaxation spectra by the Davidson-Cole equation under zero and 5T field for the bilayer system; in (b) data under 5T have been plotted using the left y-axis while that under zero field have been plotted using the right y-axis; and (d), (e), and (f) show the R$_p$, C$_p$, and $\beta$ under zero and 5T field for the perovskite systems. }
\end{figure*}

\begin{figure}[!h]
  \begin{center}
    \includegraphics[scale=.33]{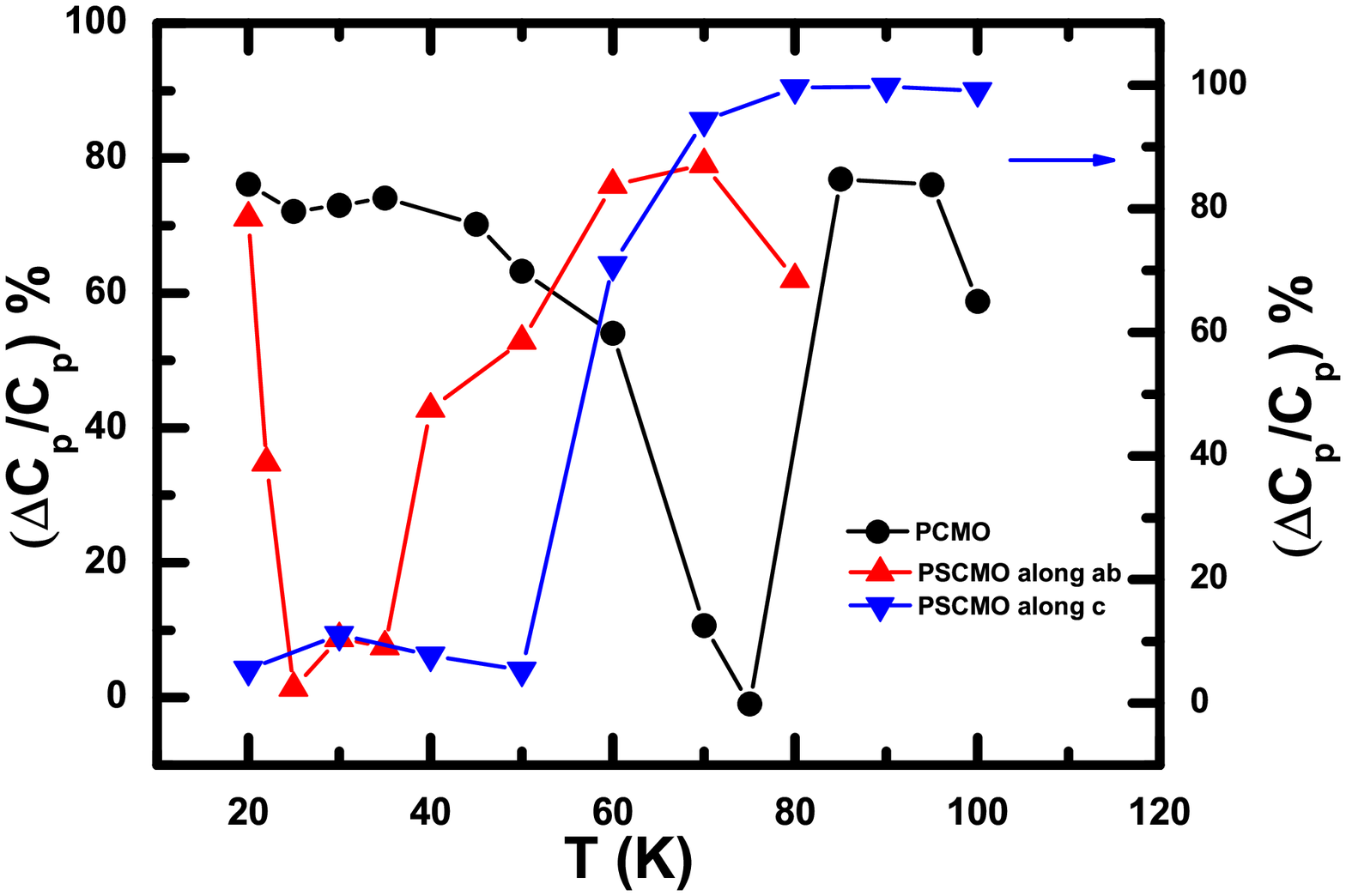} 
    \end{center}
  \caption{(color online) The magnetocapacitance for the bilayer (within and perpendicular to ab-plane) and the perovskite systems across 20-120 K. }
\end{figure}
 
We, therefore, explored the complete dielectric spectra across 100 Hz to 5 MHz. We describe the complex plane impedance plots in the $Z'-Z"$ frame,\cite{supplemental} for the entire range of temperature both under zero and 5T magnetic field, by the generalized Davidson-Cole model of relaxation\cite{Cole} 

\begin{equation}
Z^* = R_\infty + \frac{R_0 - R_\infty}{(1 + i\omega\tau_0)^\beta}
\end{equation}
where $R_0$ and $R_\infty$ are the static and high frequency resistances, respectively, $\omega$ is the frequency, $\tau_0$ is the relaxation time scale, and $\beta$ is the relaxation exponent which varies between 0 and 1 for the non-Debye correlated relaxation. Fitting the experimental data with an equivalent circuit model of a resistance ($R_p$) and a capacitance ($C_p$) connected in parallel representing the bulk resistance and the capacitance, respectively, we obtain the circuit parameters $R_p$ and $C_p$. In Fig. 4, we plot the $R_p-T$ and $C_p-T$ under zero and 5T field for both the samples. We also show the relaxation exponent $\beta$ which varies within 0.5-0.9 and signifies strongly correlated relaxation. In Figs. 4(a), (b), and (c), the $R_p-T$, $C_p-T$, and $\beta-T$ obtained along and perpendicular to the c-axis for the bilayer sample are shown. And Figs. 4(d), 4(e), and 4(f) show the corresponding plots obtained from measurement within ab-plane for the perovskite sample. 

Finally, in Fig. 5, the \% change in the circuit capacitance ($\frac{C_{5T}-C_0}{C_{5T}}$ = $\Delta C_p/C_p$) is shown for both the samples. Once again, for the bilayer sample, we show the magnetocapacitance obtained from measurements both along and perpendicular to the c-axis. The change appears to be varying within $\sim$5-90\% for both the bilayer and the perovskite systems. $\textit{This is the central result of this paper}$. The extent of anisotropy varies rather nonmonotonically within a factor of 1-5 across the entire temperature range. It is important to point out here that (i) the magnetic field influences the circuit resistance ($R_p$) as well and (ii) even though, there is a qualitative similarity between the overall temperature dependence of the magnetodielectric and magnetocapacitive effects (Figs. 3 and 5), the switch in the magnitude of the magnetodielectric effect from positive to negative or reverse could not be observed in the case of the magnetocapacitance. The circuit model applied here helps in extracting the relaxed and hence intrinsic dielectric constant of the sample. The magnetocapacitance turns out to be always positive for both the systems across 20-100 K signifying higher capacitance under a magnetic field even though the difference does have a strong temperature dependence. 

The nonmonotonic temperature dependence of the magnetocapacitance for both the systems can possibly be understood in the following way. At low temperature ($\sim$20 K), the increase in capacitance under a magnetic field could result from enhanced polarization via spin order-charge/orbital order coupling. With the increase in temperature, then, the $\Delta C_p/C_p$ decreases sharply for the bilayer and broadly for the perovskite systems. Beyond a certain temperature - varying within 40-75 K - it rises again. The decrease in $\Delta C_p/C_p$ at intermediate temperature range could result from thermalization of the charge/orbital/spin structure. Further rise at higher temperature signifies a crossover in the origin of the overall capacitance of the sample - from purely ferroelectric order to a situation where ferroelectric order coexists with charge carrier transport.\cite{Sarma,Chowdhury} The hopping of the charge carriers within a charge/spin/orbital ordered background gives rise to a stronger effect in the high temperature range. Enhanced hopping under a magnetic field as a result of local spin alignment and/or enhanced charge disproportionation - from Mn$^{3+}$, Mn$^{4+}$ to Mn$^{(3+\delta)+}$, Mn$^{(4-\delta)+}$ within an ordered structure - could possibly lead to a rise in the dielectric constant which, in turn, would yield an enhanced magnetocapacitance at higher temperature. The small charge transferred via disproportionation seems to be significant in increasing the dipole moment.\cite{Tang} The local spin alignment under a magnetic field also contributes in increasing the hopping and hence the dielectric constant. It is important to mention here that the circuit resistance ($R_p$) increases substantially in the case of the bilayer system and decreases only slightly in the case of the perovskite system. Therefore, it appears that the hopping in presence of the local charge transfer via disproportionation and spin alignment enhances the dipole moment substantially and does not contribute much to the charge conduction. It is also noteworthy, that the temperature range of the crossover varies from the bilayer to the perovskite system. This might result from variation in the extent of coupling between charge/orbital and spin degrees of freedom in different systems. Stronger coupling leads to weaker temperature dependence - as observed in the case of the perovskite system - and vice versa. We contrast our observation to that made in a non-charge/orbital ordered, non-ferroelectric system such as La$_2$NiMnO$_6$ where the asymmetric hopping in presence of disorder promotes the dielectric constant.\cite{Sarma} Application of magnetic field, in that case, reduces the disorder and hence the dielectric constant yielding negative magnetodielectric effect. In our case of a charge/orbital ordered network, on the contrary, the hopping seems to be promoted by an enhanced local alignment of spin structure (which, in turn, introduces a local disorder within an otherwise 3D antiferromagnetic background) and/or charge disproportionation. We, therefore, observe a positive magnetocapacitance effect. 

This comparsion between the magnetodielectric and magnetocapacitive effects shows that it is not sufficient to study the influence of magnetic field on the dielectric constant alone for a system where dielectric relaxation takes place over a wider frequency range and/or interface contribution too plays a significant role.\cite{Subramanian,Yasui,Tokura} One needs to extract the contribution of the purely capacitive part from an accurate description of the dielectric relaxation dynamics to find out the magnetocapacitive effect. In systems where direct determination of the ferroelectric polarization is difficult, it is important to determine the influence of magnetic field on the capacitive component of the dielectric response to estimate the intrinsic multiferroic coupling of the systems. 

\section{Summary}
In summary, we show that there is a sizable magnetocapacitance ($\sim$5-90\%) in electronic ferroelectric Pr(Sr$_{0.1}$Ca$_{0.9}$)$_2$Mn$_2$O$_7$ and Pr$_{0.55}$Ca$_{0.45}$MnO$_3$ systems below the magnetic transition point even though their ferroelectric polarization could be small. There is a remarkable difference between the magnetodielectric effect at higher frequency (1-5 MHz) and the magnetocapacitive effect even though the qualitative patterns are similar. The change-over from positive to negative or vice versa, observed in the magnetodielectric effect, could not be noticed in the magnetocapacitance part. This result shows that it is important to extract the magnetocapacitance directly for the systems where the dielectric relaxation dymanics deviates from the Debye-type relaxation by a large extent and the relaxation takes place over a wider frequency range. For the bilayer sample, both the magnetodielectric and magnetocapacitance exhibit a reasonably strong anisotropy even though the temperature dependence of them appear to follow qualitatively similar patterns. Not much anisotropy could, however, be observed for the perovskite sample. Large magnetocapacitance at higher temperature possibly originates, primarily, from strong influence of magnetic field on the hopping in presence of charge disproportionation where the role of the ferroelectric order could be smaller. Notwithstanding this, the large magnetocapacitance in these electronic ferroelectric systems, anyway, enhances the scope of their use in magnetoelectric sensor applications enormously.           
  
$\textbf{Acknowledgments.}$
This work has been supported by the Department of Science and Technology (DST), Govt of India. One of the authors' (UC) acknowledges support in the form of a Senior Research Fellowship from DST. Another author (SG) acknowledges support in the form of a Research Associateship from CSIR, Govt of India.

\end{document}